\documentclass[]{pasj00}

\begin{document}
\SetRunningHead{Asai et.al.}{Sudden Luminosity Decrease in Cir~X-1}
\Received{2014/03/14}
\Accepted{2014/05/21}

\title{Sudden End of X-Ray Outbursts around Periastron of Circinus~X-1 Observed with MAXI}

\author{
Kazumi \textsc{Asai},\altaffilmark{1}
Tatehiro \textsc{Mihara},\altaffilmark{1}
Masaru \textsc{Matsuoka},\altaffilmark{1}
Mutsumi \textsc{Sugizaki},\altaffilmark{1}
Satoshi \textsc{Nakahira},\altaffilmark{2}
Hitoshi \textsc{Negoro},\altaffilmark{3}
Motoki \textsc{Nakajima},\altaffilmark{4}
and
Atsuo T. \textsc{Okazaki}\altaffilmark{5}
}
\altaffiltext{1}{MAXI team, RIKEN, 2-1 Hirosawa, Wako, Saitama 351-0198}
\email{kazumi@crab.riken.jp}
\altaffiltext{2}{ISS Science Project Office, ISAS, JAXA, 2-1-1 Sengen,
Tsukuba, Ibaraki 305-8505}
\altaffiltext{3}{Department of Physics, Nihon University,
1-8-14 Kanda-Surugadai, Chiyoda-ku, Tokyo 101-8308}
\altaffiltext{4}{School of Dentistry at Matsudo, Nihon University, 
2-870-1 Sakaecho-nishi, Matsudo, Chiba 271-8587}
\altaffiltext{5}{Faculty of Engineering, Hokkai-Gakuen University,
Toyohira-ku, Sapporo 062-8605}

\maketitle

\begin{abstract}

MAXI/GSC observed 21 outbursts from Circinus~X-1
between 2009 August and
\textcolor{black}{2013 December}. 
Although 14 outbursts showed ordinary gradual decays,
in 7 outbursts we found sudden luminosity decrease
\textcolor{black}{
in a time scale of a few hours
around the periastron, and then the outbursts terminated.}
These sudden decreases started at the estimated luminosity of
a few times $10^{37}$ erg s$^{-1}$ and reached to
$\lesssim3\times10^{36}$ erg s$^{-1}$.
\textcolor{black}{
We propose three interpretations}
for the sudden luminosity decrease:
(1) the end of the outburst during the dip,
(2) the propeller effect, and 
(3) the stripping effect by the stellar wind of the companion star.
It is difficult to explain the phenomenon with any of these
\textcolor{black}{interpretations} alone.
\textcolor{black}{
The interpretation of (1) is possible for only two outbursts
assuming rapid decay.}
The propeller effect (2) is expected to occur at a constant luminosity,
which is incompatible with the observed facts. 
In wind stripping \textcolor{black}{effect} (3),
the ram pressure of a typical stellar wind is not sufficient to blow out
most of the accretion disk.
\textcolor{black}{
In this paper, we discuss a possibility of a modified effect
of (3) assuming other additional conditions such as wind clumping
and disk instability.
}
\end{abstract}

\section{Introduction}

Circinus X-1 (hereafter Cir~X-1) is a bright X-ray binary located on
the Galactic plane.
The nature of the large variability in various time scales
has not been well understood yet.
The source shows a recurrent X-ray activity coupled with
an orbital period of 16.6~d \citep{Kaluzienski1976}.
It is also known to show long-term variabilities
on a time-scale of years, and the flux level varies 
from undetectable levels to more than 1~Crab
(\cite{Parkinson2003}; \cite{Clarkson2004}; \cite{Armstrong2013}).
When the luminosity is high,
it frequently showed X-ray outbursts and radio flares by the 16.6~d
orbital cycle.
The flaring events are believed to be caused by 
an increased mass transfer from the companion star near the periastron passage
of a highly eccentric binary orbit
(\cite{Kaluzienski1976}; \cite{Jonker2007}).
The X-ray outbursts are accompanied by absorption
dips in the X-ray flux.
\citet{Clarkson2004} reported that the X-ray dips would provide more
accurate system clock than the maximum of the X-ray outbursts.
Recently, periodic radio flares are considered more suitable for
defining the orbital motion, and used as the phase zero in the ephemeris 
\citep{Nicolson2007}.

X-ray spectra observed with ASCA, RXTE, and BeppoSAX during the dips
are consistent with a partial covering model of absorbing matter in
which the intensity of the intrinsic emission does not change 
(\cite{Brandt1996}; \cite{Shirey1999b}; 
\textcolor{black}{
Iaria et al. 2001a, 2001b).
}  
Observations of Chandra HETGS near periastron have revealed presence
of emission lines of H-like and/or He-like Ne, Mg, Si, S, and Fe.
These lines are broad and show P Cygni profiles with velocities of
$\pm\,2000$~km s$^{-1}$ \citep{Brandt2000}.
These results suggest that a wind with a moderate
temperature ($\sim 5\times10^6$~K)
from the X-ray-heated accretion disk is responsible for the P Cygni profiles.

Cir~X-1 does not fit the conventional classification of X-ray binaries,
either HMXBs (high-mass X-ray binaries) or LMXBs (low-mass
X-ray binaries) \citep{White1989}. 
\textcolor{black}{
In addition to type-I X-ray bursts (\cite{Tennant1986}; \cite{Linares2010}),
spectral and timing features revealing both Z-source and atoll-source
(\cite{Shirey1999a}; \cite{Soleri2009})
and 
rapid variability with kHz quasi-periodic oscillations
\citep{Boutloukos2006}}
agree with a typical behaviors
in LMXBs embedding a weakly magnetized neutron star ($B\sim 10^8-10^9$~G).
\textcolor{black}{
However, results of optical spectroscopic and photometric
observations suggest that the companion star is a 
B5--A0 supergiant 
and has an eccentric binary orbit of $e=0.45$ \citep{Jonker2007}.
Those properties indicate a HMXB, which usually accompanies
a strongly magnetized neutron star
($B\sim 10^{11}-10^{12}$~G).
The system is really young as \citet{Heinz2013} discovered
a young ($< 4600$~yr)
X-ray supernova remnant surrounding Cir~X-1.
To explain the discrepancy, \citet{Heinz2013} suggested
either that the neutron star magnetic field
decayed quickly to below $10^{12}$~G or that the neutron star
was born with such a low field.
There was also the same suggestion in \citet{Jonker2007}.
}

The long-term X-ray activity had been continuously monitored with RXTE
\citep{Bradt1993} ASM (All Sky Monitor: \cite{Levine1996})
since 1996 to 2011 December.
Until 2000, the source had been in a very active phase with a daily average
flux of $\sim$1~Crab \citep{Parkinson2003}.
Since 2000, the activity had declined gradually, and then
it reached to the inactive phase with a flux below the detection limit
of 10 mCrab and no outburst activity in 2007.
In 2009 August, MAXI (Monitor of All-sky X-ray Image: \cite{Matsuoka2009})
started the all-sky survey with GSC
(Gas Slit Camera: \cite{Mihara2011}; \cite{Sugizaki2011}) 
and SSC (Solid-state Slit Camera: \cite{Tomida2011})
on the International Space Station.
The MAXI nova alert system \citep{Negoro2010}
detected the beginning of an outburst from Cir~X-1
on 2010 May 7 \citep{Nakajima2010}.
The X-ray flux of the outburst reached
$320\pm30$~mCrab on May 8. 
After the event, the source return to the active phase. 
\citet{DAi2012} analyzed data of RXTE, Swift, and Chandra observations
of outbursts from 2010 May 11 to July 4, and
reported that the source exhibited a clear spectral transition
from an optically thick state (soft state) to
an optically thin state (hard state) which can be interpreted
within the disk-instability scenario.

In this paper, we report the recent outburst properties of Cir~X-1
obtained from MAXI GSC light curve from 2009 August 15 to
\textcolor{black}{
2013 December 18.
}
In particular, we investigate the observed behavior of sudden
luminosity decrease \textcolor{black}{around} the periastron passages.
The paper is organized as follows.
In section~2, we describe the observations and
the light curve analysis, and then estimate the radius of the outer disk.
In section~3, we discuss three \textcolor{black}{interpretations} for the
\textcolor{black}{
new}
sudden luminosity decrease. 
Finally, we present the conclusions in section~4.

\section{
 \textcolor{black}{Observational Data and Analysis}}

 \subsection{X-ray Light Curves of Outbursts}

 \textcolor{black}{
In this section we examine the behavior of outbursts of Cir~X-1 using
the light curve data observed by
MAXI/GSC\footnote{$<$http://maxi.riken.jp/$>$.}
and by
Swift \citep{Gehrels2004}/BAT (Burst Alert Telescope: \cite{Barthelmy2005}).
\footnote{$<$http://heasarc.gsfc.nasa.gov/docs/swift/results\\/transients/$>$.}
The comparison of both the data is useful to investigate soft/hard state
and propeller effect of neutron star LMXBs similar to Cir~X-1 
(Asai et al. 2012, 2013).
To make easy to compare both data, the count rates of GSC and BAT
are converted to luminosities by assuming
Crab-like spectrum \citep{Kirsch2005} and by using a distance of 
$7.8$~kpc employed by \citet{DAi2012}.
Furthermore, because Cir~X-1 has a 16.6~d binary period we employ
the orbital cycle number as a time scale.
The ephemeris is employed from \citet{Nicolson2007} as was in \citet{DAi2012},
which is expressed as}
\begin{eqnarray}
T_{\rm MJD}\ (N) = 43076.32 + (16.57794\pm0.00094) N \nonumber \\
- (0.0000401\pm0.0000015)N^2,
\end{eqnarray}
where $N$ is the number of 16.6~d cycles since MJD = 43076.32
and refer to the orbital phase calculations based on this ephemeris.
Here, we also adopt that the phase 0 corresponds to the periastron
because this ephemeris predicts start of most of
the X-ray outbursts as mentioned in \citet{DAi2012}. 

\begin{figure*}
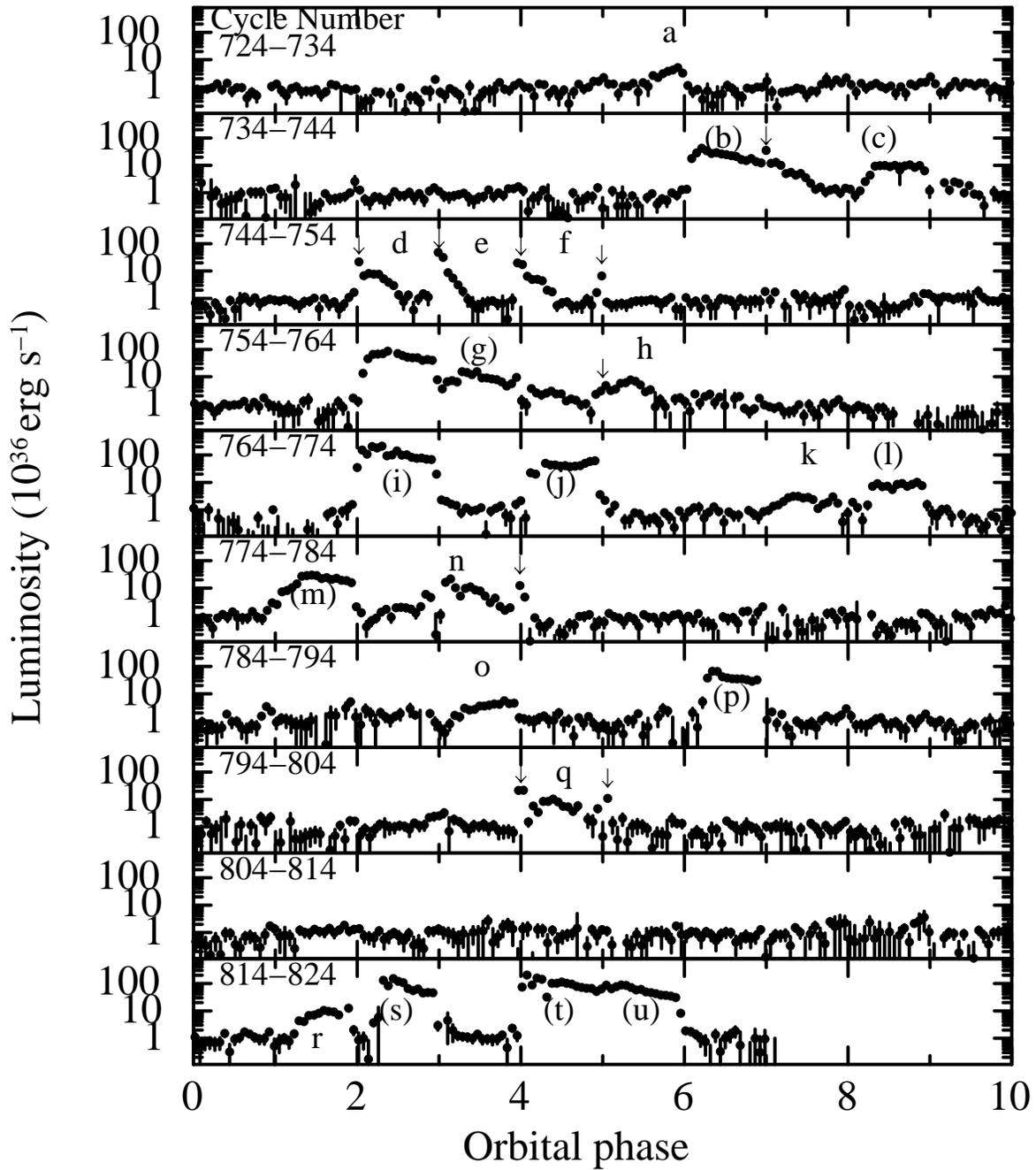

  \begin{center}
    \FigureFile(160mm,){fig1-new.eps}
  \end{center}
  \caption{
Light curve in 2-10~keV luminosity (d=7.8~kpc) against orbital phase.
The data are one-day averaged.
\textcolor{black}{
The alphabets (a--u) denote the outbursts observed.
The alphabets with parentheses denote that the outbursts continue
until next periastron.
}
Arrows indicate the data points of ``periastron spike''(see the text).
}
\label{fig1}
\end{figure*}

\begin{figure*}
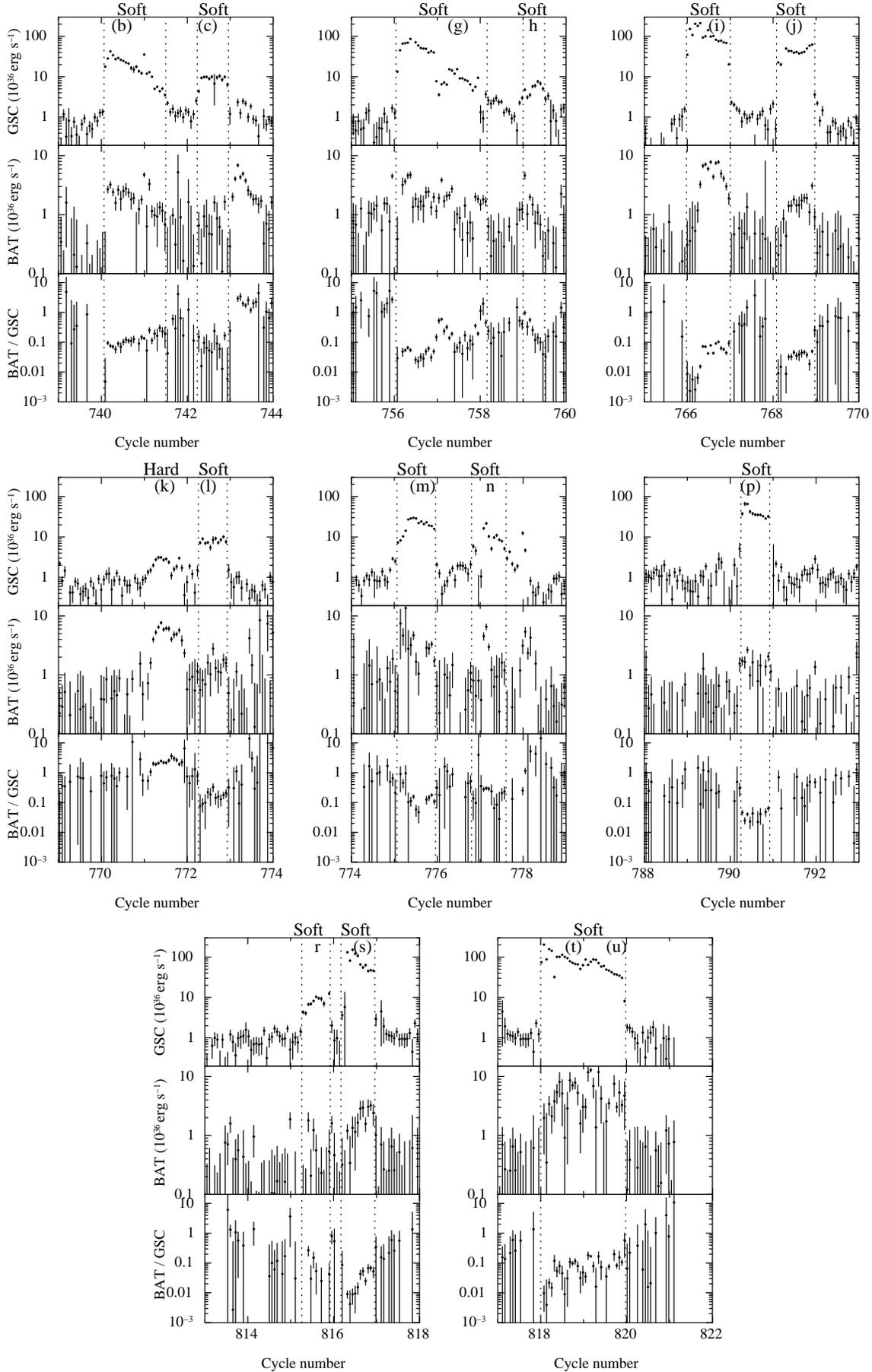

  \begin{center}
    \FigureFile(50mm,){fig2-1-new.eps}
    \FigureFile(50mm,){fig2-2-new.eps}
    \FigureFile(50mm,){fig2-3-new.eps}
    \FigureFile(50mm,){fig2-4-new.eps}
    \FigureFile(50mm,){fig2-5-new.eps}
    \FigureFile(50mm,){fig2-6-new.eps}
    \FigureFile(50mm,){fig2-7-new.eps}
    \FigureFile(50mm,){fig2-8-new.eps}
  \end{center}
  \caption{
One-day GSC light curves (2--10~keV), one-day BAT light curves (15--50~keV),
and the hardness ratios (BAT/GSC)
\textcolor{black}{
for the outburst periods remarked in figure~\ref{fig1}.}
The region between two vertical dotted lines indicates the soft state
\textcolor{black}{
as noted above the figure}.
The vertical error bars represent 1-$\sigma$ statistical uncertainty.
}
\label{fig2}
\end{figure*}

\textcolor{black}{
Figure~\ref{fig1} shows the one-day averaged GSC light curve
in 2--10~keV luminosity against the orbital phase.
The light curve starts from MJD = 55058.5 (2009 August 15) until
MJD = 56644.5 (2013 December 18).
There were 21 outbursts denoted as alphabet (a--u) in the figure.
Here we chose remarkable eleven outbursts which continue
until the next periastron.
These are indicated by alphabets with parentheses in the figure.}

\textcolor{black}{
These remarkable outbursts are shown in figure~\ref{fig2},
as the one-day GSC light curves (2--10~keV),
one-day BAT light curves (15--50~keV), and the hardness ratios (BAT/GSC).}
In the GSC light curve, the detection limit is
$\sim 1\times 10^{36}$ erg s$^{-1}$.
However, the detection limit sometimes increases
because of the variability of the background.
For example, between outbursts (i) and (j), we cannot recognize
the source in the daily GSC image,
although the average luminosity was estimated as
$(1.5\pm1.0)\times10^{36}$ erg s$^{-1}$.
The spectral state was \textcolor{black}{mostly}
determined by the BAT(15--50~keV)/GSC(2--10~keV)
hardness ratio, where a ratio less than $\sim1$ identifies the soft state
\textcolor{black}
{according to}
the method employed in \citet{Asai2012}.
However, when the data points for the GSC and/or BAT  have large errors,
it is difficult to distinguish
\textcolor{black}{
the states.
}
In this case, we judged the state using the luminosity of
the spectral state transition
\textcolor{black}{
of the outburst (b)} reported by \citet{DAi2012}, that is,
$(3.5\pm0.7)\times10^{36}$ erg s$^{-1}$.
\textcolor{black}{Thus obtained} periods of the soft state
are denoted in figure~\ref{fig2}.

\begin{figure}
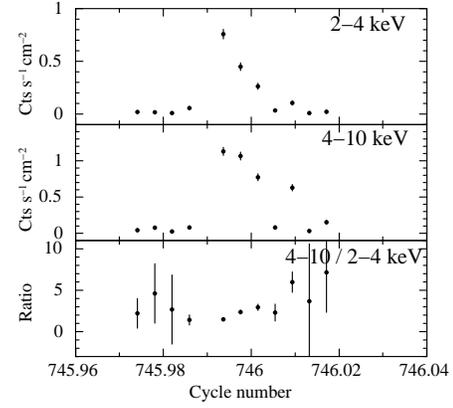

  \begin{center}
    \FigureFile(60mm,){fig3-new.eps}
  \end{center}
  \caption{
GSC light curves and hardness ratio of periastron spike at cycle number = 746.
The time of the spike is MJD = 55421.0432.
Data points are of each scan.
Decay of flux is seen over several scans in this spike. 
The vertical error bars represent 1-$\sigma$ statistical uncertainty.
}
\label{fig3}
\end{figure}

\textcolor{black}{
We note the following four points
from figures~\ref{fig1} and \ref{fig2} :
}
\begin{description}
\item[(1)]
\textcolor{black}{
The seven outbursts, (i), (j), (l), (m), (p), (s), and (u),
reveal sudden luminosity decrease around the periastron, and then
the outbursts terminated. 
Such decreasing feature has not been detected in any outburst
of LMXBs embedding weakly magnetized neutron star.
This feature is a main subject to be investigated in this paper.
}
\item[(2)]
There are three outbursts (i), (s), and (t), which show very soft state
(\textcolor{black}{BAT/GSC} $\sim 0.01$).
Their peak luminosities in the 2--10~keV band (GSC)
are very high,
\hspace{0.3em}\raisebox{0.4ex}{$>$}\hspace{-0.75em}\raisebox{-.7ex}{$\sim$}\hspace{0.3em}
$1.5\times10^{38}$ erg s$^{-1}$, which is almost the Eddington luminosity.
On the other hand, the luminosities in the 15--50~keV band (BAT) are very
low
(\hspace{0.3em}\raisebox{0.4ex}{$<$}\hspace{-0.75em}\raisebox{-.7ex}{$\sim$}\hspace{0.3em}
$1\times10^{36}$ erg s$^{-1}$).
This means that the blackbody emission increases  without 
Comptonized emission at the onset.
This very soft state would be another disk state, such as a slim disk,
which is sometimes observed in Z sources
(e.g., \cite{Lin2012}).
\item[(3)]
\textcolor{black}{
At the peak of outburst k, the BAT/GSC hardness ratio remained above 1.
This indicates the hard state throughout the outburst.
Although the peak of the other outbursts is usually in the soft state,
this is only one exception in our observation (see figure~\ref{fig2}).
The peak luminosity of this outburst in 2--10~keV 
was low ($\sim 3\times10^{36}$ erg s$^{-1}$),
whereas that in 15--50~keV was high (i.e., a ``hard outburst'').
}
\item[(4)]
Jumps (\textcolor{black}{burst-like} points)
appeared between phases $-0.05$ and 0.05
at nine periastron passages,
indicated by arrows in figure~\ref{fig1}.
The luminosities are above $1\times10^{37}$ erg s$^{-1}$ and
they occurred independently of the outbursts.
We call them ``periastron spikes.''
The properties of the nine periastron spikes are summarized
in \textcolor{black}{table~\ref{tab1}}.
\textcolor{black}{
Here, we used the data of each scan,
which is usually performed once during the 92~min orbital period of MAXI. 
The
}
detailed analysis revealed that the hardness ratio
of 4--10~keV to 2--4~keV did not show any softening,
as an example is shown in figure~\ref{fig3}.
\textcolor{black}{Thus, the} periastron spikes are not type-I X-ray bursts.
It would be caused by a temporary increase of the accretion rate
independent of the existence of the accretion disk, 
such as spherical accretion from the polar directions.
\textcolor{black}{
Further discussion on this spike will be made in subsection~3.3}
\end{description}

\begin{table}
\caption{Periastron spike.}
\label{tab1}
\begin{center}
\begin{tabular}{cccc}
\hline
Cycle & Peak & Peak & Location \\ 
Number\footnotemark[$*$] & phase & Luminosity\footnotemark[$\dagger$]  
& to an outburst \\
\hline
741 &   0.0023  & $49\pm3$ & during \textcolor{black}{outburst} (b)\\
746 & $-0.0071$ & $93\pm4$ & onset of \textcolor{black}{outburst} d\\
747 & $-0.0005$ & $161\pm6$ & onset of \textcolor{black}{outburst} e\\
748 & $-0.0130$ & $173\pm7$ & onset of \textcolor{black}{outburst} f\\
749 & $-0.0330$ & $21\pm2$ & no outburst \\
759 &   0.0170  & $12\pm2$ & during \textcolor{black}{outburst} h \\
778 &   0.0171  & $24\pm3$ & end of \textcolor{black}{outburst} n \\
798 &   0.0031  & $250\pm10$ & onset of \textcolor{black}{outburst} q\\
799 &   0.0363  & $17\pm3$ & end of \textcolor{black}{outburst} q  \\
\hline
\multicolumn{3}{@{}l@{}}{\hbox to 0pt{\parbox{80mm}{\footnotesize
\par\noindent
\footnotemark[$*$]
Orbital cycle number defined  by \citet{Nicolson2007}. 
\par\noindent
\footnotemark[$\dagger$]
Luminosity in unit of $10^{36}$ erg s$^{-1}$ in 2--10~keV energy band. 
The errors are in 1-$\sigma$.
}\hss}}
\end{tabular}
\end{center}
\end{table}

\subsection{Variation \textcolor{black}{around} Periastron Passage}

\begin{table*}
\caption{Outbursts observed by MAXI/GSC.}
\label{tab2}
\begin{center}
\begin{tabular}{cccccccc}
\hline
Outburst\footnotemark[$*$]  & \textcolor{black}{Onset}\footnotemark[$\dagger$]
& Peak day\footnotemark[$\ddagger$]
& Peak & State & Behaviour & Luminosity\footnotemark[$\S$] & Luminosity\footnotemark[$\S$] \\ 
          & (phase) & (MJD) & luminosity\footnotemark[$\S$] & at peak &
\textcolor{black}{\small around} 
\small periastron\footnotemark[$\l$]
& \small{before sudden drop}
& \small{after sudden drop}\\
\hline
a & $-0.41\pm0.03$ & 55155.5 & $5.1\pm0.3$  & soft & not clear & -- & -- \\
(b) & $0.06\pm0.03$ & 55325.5 & $42.4\pm0.9$ & soft & continuous
& ($12\pm1$)\footnotemark[$\#$] & -- \\
(c) & $0.18\pm0.03$ & 55362.5 & $9.9\pm0.6$ & soft & continuous
& ($6.3\pm0.7$)\footnotemark[$\#$] & -- \\
d & $-0.07\pm0.03$ & 55423.5 & $8.1\pm0.3$ & soft & -- & -- & -- \\
e & $-0.07\pm0.06$ & 55439.5 & $8.7\pm0.7$ & soft & -- & -- & -- \\
f & $-0.07\pm0.03$ & 55455.5 & $6.5\pm0.3$ & soft & -- & -- & -- \\
(g) & $0.04\pm0.03$ & 55592.5 & $86\pm2$ & soft & continuous
& ($39.9\pm0.7$)\footnotemark[$\#$] & -- \\
h & $-0.12\pm0.03$ & 55641.5 & $7.6\pm0.6$ & soft & -- & -- & -- \\
(i) & $-0.03\pm0.03$ & 55752.5 & $153\pm2$ & soft & sudden drop & $68.4\pm0.8$ & $1.5\pm1.0$ \\
(j) & $0.08\pm0.03$ & 55799.5 & $62\pm2$ & soft & sudden drop & $62\pm2$ & $< 1.6$ \\
k & $0.05\pm0.03$ & 55840.5 & $3.1\pm0.2$ & hard & -- & -- & -- \\
(l) & $0.20\pm0.03$ & 55864.5 & $10.1\pm0.8$ & soft & sudden drop & $7.6\pm0.8$ & $< 2$ \\
(m) & $-0.14\pm0.03$ & 55907.5 & $29.8\pm 0.6$ & soft & sudden drop & $15.8\pm0.4$ & $1.5\pm1.0$ \\
n & $-0.26\pm0.03$ & 55935.5 & $22\pm1$ & soft & -- & -- & -- \\
o & $0.15\pm0.03$ & 56111.5 & $5.5\pm0.5$ & soft & not clear & -- & -- \\
(p) & $0.18\pm0.03$ & 56153.5 & $66\pm6$ & soft & sudden drop & $32\pm2$ & $1.7\pm1.5$ \\
q & $-0.01\pm0.09$ & 56286.5 & $10\pm1$ & soft & -- & -- & -- \\
r & $0.24\pm0.03$ & 56575.5 & $13\pm1$ & soft & -- & -- & -- \\
(s) & $0.12\pm0.06$ & 56584.5 & $152\pm10$ & soft & sudden drop & $46\pm3$ & $1.7\pm1.4$ \\
(t) & $-0.05\pm0.03$ &  56611.5 & $203\pm3$ & soft & next outburst & 
($62\pm2$)\footnotemark[$\#$] & -- \\
(u) & $-0.07\pm0.03$ & 56630.5 & $88\pm1$ & soft & sudden drop & $30.6\pm0.9$ & $1.5\pm1.3$ \\
\hline
\multicolumn{8}{@{}l@{}}{\hbox to 0pt{\parbox{180mm}{\footnotesize
\par\noindent
\footnotemark[$*$]
\textcolor{black}{
The alphabets with parentheses denote that the outbursts continue
until next periastron.}
\par\noindent
\footnotemark[$\dagger$]
In order to define the ``onset phase'', we defined ``last off state''
and ``first on state''.
The ``last off state'' indicates the lowest data point, 
after which the flux increased monotonically 
towards the peak of the outburst.
The ``first on state'' indicates the first data point which has a
significantly higher flux than that of the ``last off state''.
The ``onset phase'' is defined by the middle of the ``last off state''
and the ``first on state'', and the error is defined by the span between
the both data points.
\par\noindent
\footnotemark[$\ddagger$]
Day of maximum luminosity in the 2--10~keV GSC light curve in one day bin. 
\par\noindent
\footnotemark[$\S$]
Luminosity in unit of $10^{36}$ erg s$^{-1}$ in 2--10~keV energy band.
The errors are 1-$\sigma$ errors.
\par\noindent
\footnotemark[$\l$]
The X-ray behaviour \textcolor{black}{around} periastron passage.
``sudden drop'' denotes that the luminosity suddenly decreases 
\textcolor{black}{around} periastron passage and the outburst seems
to have ended.
``continuous'' 
\textcolor{black}{(continuous outburst)}
denotes that the outburst continues after periastron passage.
``next outburst'' 
\textcolor{black}{(next new outburst)}
denotes that next outburst seems to start
after periastron passage.
``not clear'' denotes that the distinction of ``sudden drop''
and ``dip'' is not clear, because of low peak luminosity.  
``--'' denotes that the outburst had ended before next periastron passage. 
\par\noindent
\footnotemark[$\#$]
Luminosity at phase \textcolor{black}{$\sim 0.95$} before next dip.
}\hss}}
\end{tabular}
\end{center}
\end{table*}

\begin{figure}
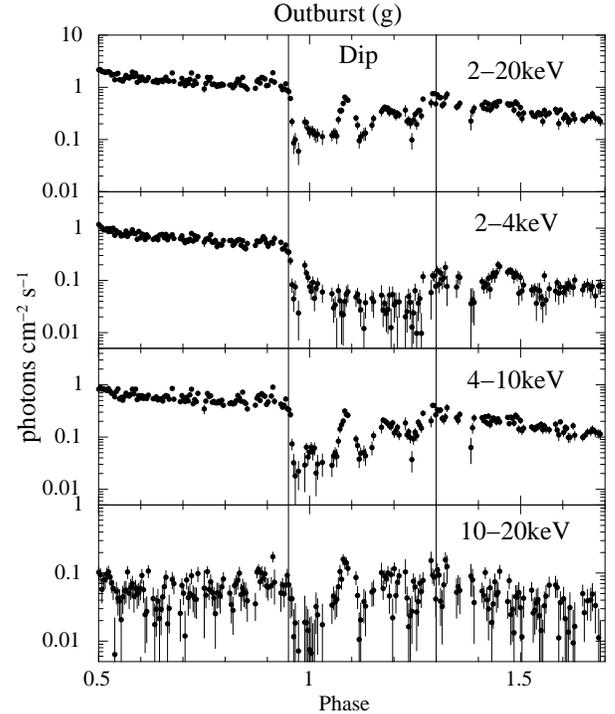

  \begin{center}
    \FigureFile(80mm,){fig4-new.eps}
  \end{center}
  \caption{
    \textcolor{black}{
One-orbit (92-min) averages GSC light curves of outburst (g)
in four energy bands around the next
periastron passage (phase 1 is $N = 757$).
The dip was seen from phase $\sim 0.95$ to $\sim 1.3$, which
are shown by the region between two vertical lines. 
}
The vertical error bars represent 1-$\sigma$ statistical uncertainty.
}
\label{fig4}
\end{figure}

\begin{figure*}
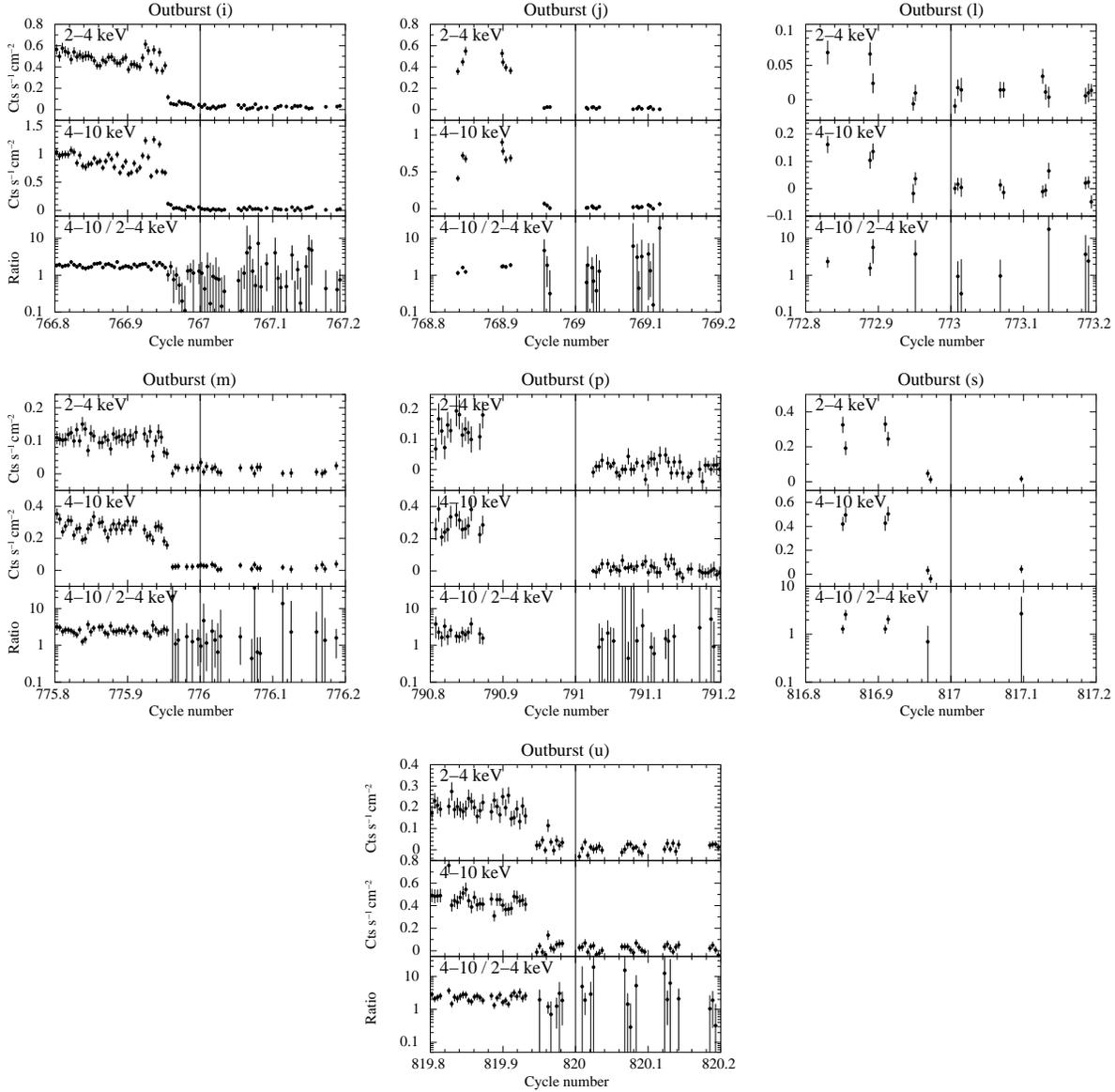

  \begin{center}
    \FigureFile(52mm,){fig5-1-new.eps}
    \FigureFile(52mm,){fig5-2-new.eps}
    \FigureFile(52mm,){fig5-3-new.eps}
    \FigureFile(52mm,){fig5-4-new.eps}
    \FigureFile(52mm,){fig5-5-new.eps}
    \FigureFile(52mm,){fig5-6-new.eps}
    \FigureFile(52mm,){fig5-7-new.eps}
  \end{center}
  \caption{
Magnified GSC light curves of outbursts and hardness ratios around periastron
passage \textcolor{black}{for seven outbursts
classified as``sudden drop'' in subsection~2.2.
The vertical line indicate a phase of periastron.}
Data points are of each scan.
The vertical error bars represent 1-$\sigma$ statistical uncertainty.
In outburst (p), there is a large data gap around the periastron
(cycle number 791)}
\label{fig5}
\end{figure*}

\textcolor{black}{
The properties of 21 outbursts (a--u) are summarized in table~\ref{tab2}.
As shown in the table, the outbursts start in an orbital phase between 
$\sim -0.3$ and $\sim$ 0.2}, except outburst a. 
Eleven outbursts
\textcolor{black}{
(alphabets with parentheses in figure~\ref{fig1} and ~\ref{fig2}, and
remarkable outbursts in subsection 2.1)}
continued to the next periastron.
The behavior at the next periastron can be classified into three types:
``continuous outburst''
\textcolor{black}{[i.e., (b), (c), and (g)],}
 ``sudden drop'' 
\textcolor{black}{[i.e., (i), (j), (l), (m), (p), (s), and (u)],
``next new outburst''[i.e., (t)]},
which are denoted in table~\ref{tab2}.
The features of three types are summarized as below: 
\begin{description}
\item[\textcolor{black}{Continuous outbursts:}]
Outbursts (b), (c), and (g) continued smoothly to
the next orbital cycle
after periastron passage, although the intensity decreased temporarily
around periastron.
The intensity decreases are interpreted as the absorption dip
that \textcolor{black}{was} often observed before an outburst onset
when the luminosity was high in 1996--2000 \citep{Clarkson2004}.
\textcolor{black}{
In particular, the outburst (g) shows a clear absorption dip, which
can be understood from the light curves divided in energy band.
In figure~\ref{fig4}, we show the light curves of outburst (g)
in four energy bands, 2--20~keV, 2--4~keV, 4--10~keV, and 10--20~keV,
using one-orbit (92-min) averaged data.}
The energy dependence of the light curves shows that
the dip is caused by a cold absorber.
\textcolor{black}{
The dip continued from phase \textcolor{black}{$\sim 0.95$ to $\sim 1.3$.}
In outbursts (b) and (c), the dip is not clear in the GSC light curve.
However, \citet{DAi2012} reported that 
the spectral shape was affected by the variable cold absorber.
They obtained the absorption values of 
$N_{\rm H} \sim 5\times10^{22}$~cm$^{-2}$ and 
$N_{\rm H} \sim 6\times10^{23}$~cm$^{-2}$ for
outburst (b) (around cycle number 741) using RXTE data and
outburst (c) (around cycle number 743) using Swift data,
respectively.
}

\citet{DAi2012} also reported that
the outburst (c)
seems to turn off suddenly \textcolor{black}{near} periastron passage (cycle number 743)
in the ASM light curve.
In the GSC light curve (figure~\ref{fig2}), the 
outburst seems to continue with a gradual decrease, although
there is a large data gap.
The BAT light curve shows a sudden increase
just after periastron passage, which is like a hard outburst.
In the present data, we cannot distinguish
either whether outburst (c) ended and another hard outburst started,
or whether the state changed to a hard state \textcolor{black}{during}
outburst (c). 

\item[\textcolor{black}{Next new outburst:}]

Outburst (t) also appears to continue to the next orbital cycle after
periastron passage.
However, the decay profile does not continue to that after dip
(see figure~\ref{fig6} in subsection 2.3).
A new outburst seems to occur at the next periastron during the soft state.
This behavior resembles that of the outbursts caused by the enhanced
mass transfer when the luminosity was high
in 1996--2000
\textcolor{black}{
(e.g., \cite{Clarkson2004}).
}
We classified outburst (t) as the ``next \textcolor{black}{new} outburst''.

\item[\textcolor{black}{Sudden drop:}]
\textcolor{black}{
In contrast to the continuous outburst and the next
\textcolor{black}{new} outburst,
outbursts (i), (j), (l), (m), (p), (s), and (u) show a sudden 
luminosity decrease around periastron.
\textcolor{black}{
The seven outbursts were also described in subsection 2.1 (1).
} 
In order to examine the detailed variability around
the sudden decrease phase, we plotted 
their detailed GSC light curves and
GSC hardness ratios (4--10 / 2--4 ~keV) using the short time
scale data of each scan (92~min) as shown in figure~\ref{fig5}.
The sudden decrease at phase $\sim 0.95$ is seen clearly
in (i), (m), and (u), in which the time scale is about 90~min.
The sudden luminosity decreases of other outbursts (j), (l), (p),
and (s) are consistent with the occurrence at a phase around
0.95 although there are considerable data gaps.
}

The sudden luminosity decreases at phase
\textcolor{black}{$\sim 0.95$}
resembles the
dip of outburst (g) \textcolor{black}{
in the point of the time-scale of a few hours.}
However, in these seven outbursts, the luminosities
\textcolor{black}{after the sudden drop} remained low
\textcolor{black}{in hard state}
until the next outburst.
The luminosities before and after the sudden drop are summarized
in table~\ref{tab2}.
The luminosity range just before the sudden drop
is (6--70) $\times 10^{36}$ erg s$^{-1}$, 
\textcolor{black}{
and
}
the spectral state is \textcolor{black}{in soft state}
judging from the hardness ratio (BAT/GSC) in figure~\ref{fig2}.
These results are consistent with the report of
\citet{DAi2012} that the luminosity of the spectral state transition occurs at
$(3.5\pm0.7)\times10^{36}$ erg s$^{-1}$.
After the sudden drop, the luminosities stay below
$\sim 3\times10^{36}$ erg s$^{-1}$.
Although the spectral state is not clear from the hardness
ratio (BAT/GSC) in figure~\ref{fig2} because of large uncertainties,
the low luminosities indicate that it is most likely
the hard state \citep{DAi2012}.
We discuss the possible cause of the sudden drop in the next section.

\textcolor{black}{
Here, it is noted that there is an interesting information on radio
observations for outburst (m) although radio observations for
others are not reported.
Namely,}
large radio flares were observed after the periastron passages
of the end of outburst (m) (cycle number = 776)
and the beginning of outburst n (cycle number = 777)
\citep{Armstrong2013}.
\end{description}

\subsection{Estimate of Outer Radius from Outburst Decay Curve}
\label{outerradius}
In the MAXI \textcolor{black}{Cir~X-1} observations,
the persistent luminosity was low and
the source sometimes caused an X-ray outburst.
This behavior reminds us of soft X-ray transients
\citep{Campana1998a}.
Their recurrent outburst behavior
is usually interpreted by the 
thermal viscous disk instability model (DIM: \cite{Lasota2001}).
\citet{DAi2012} reported 
\textcolor{black}{to have analyzed outburst (b)}
that the source showed a clear spectral state transition, 
from an optically thick state (soft state)
to an optically thin state (hard state)
and
\textcolor{black}{
that this is}
interpreted it via the DIM.
Therefore, we also analyze the light curves of the outbursts
on the basis of the DIM.

The morphology of the outbursts is generally characterized by the
fast rise exponential decay (FRED) behavior, and
the DIM is essentially able to explain it.
However, the DIM alone cannot reproduce a realistic FRED outburst.
The X-ray irradiation effect is considered to be very important.
If the edge of the outer disk is sufficiently ionized by irradiation,
outburst exponentially decays \citep{King1998}.
In contrast, if the disk irradiation is not strong enough,
the edge of the outer disk is neutral and linear decay is expected.
This also means that the exponential decay may change to the linear decay
when the X-ray luminosity decreases sufficiently during the decay phase.
Powell, Haswell, and Falanga (2007) labeled this change in the light curve
as a ``brink.''
In this subsection, we estimate the outer radius of the accretion disk
by using 
\textcolor{black}{
the fitting results of the linear and exponential function
}
under the DIM model.
\textcolor{black}{
In this estimation},
we used the one-orbit (92-min) average GSC light curve for the
fitting.

\begin{figure*}
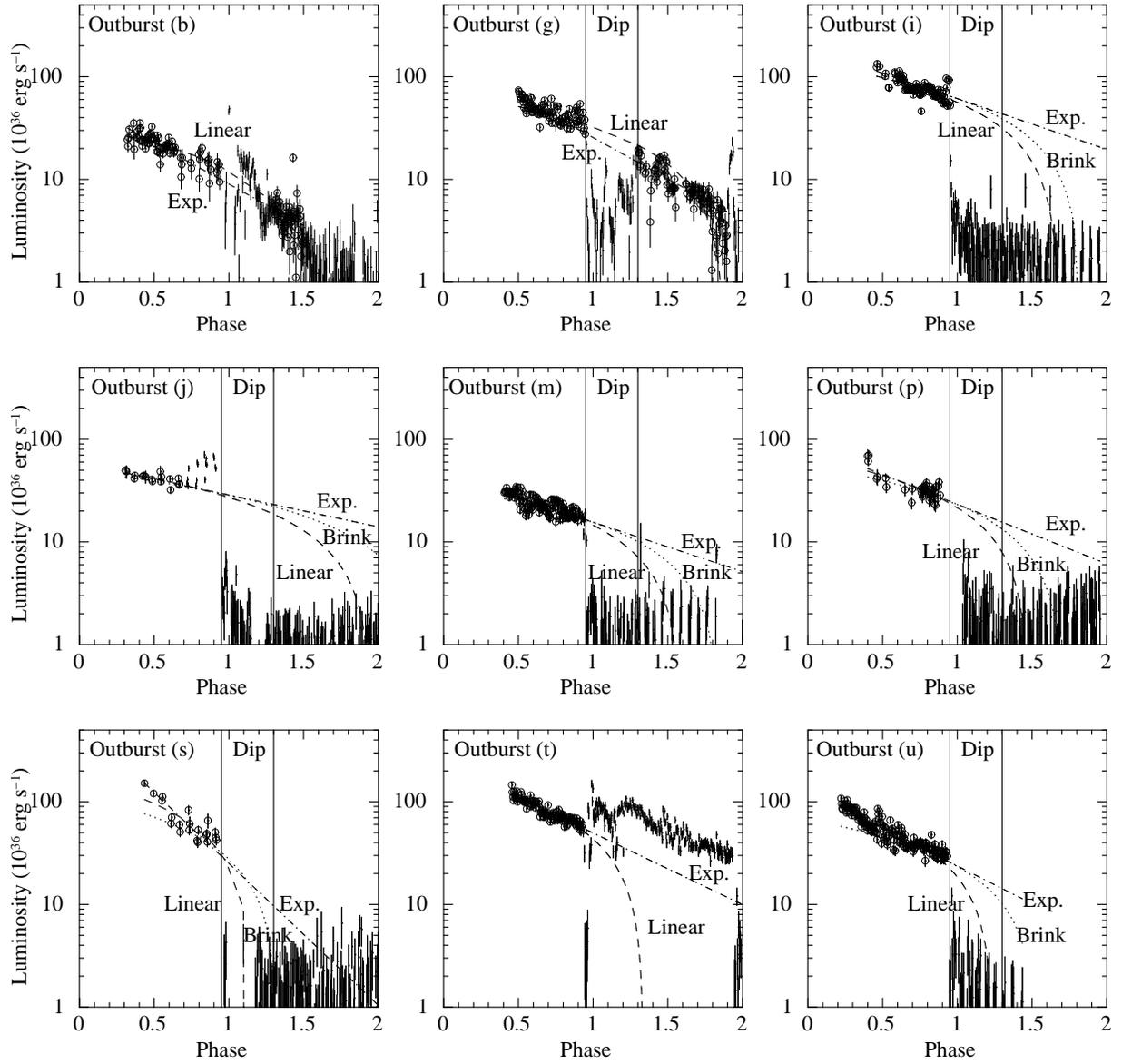

  \begin{center}
    \FigureFile(56mm,){fig6-1-new.eps}
\hspace{-5mm}
    \FigureFile(56mm,){fig6-2-new.eps}
\hspace{-5mm}
    \FigureFile(56mm,){fig6-3-new.eps}
    \FigureFile(56mm,){fig6-4-new.eps}
\hspace{-5mm}
    \FigureFile(56mm,){fig6-5-new.eps}
\hspace{-5mm}
    \FigureFile(56mm,){fig6-6-new.eps}
    \FigureFile(56mm,){fig6-7-new.eps}
\hspace{-5mm}
    \FigureFile(56mm,){fig6-8-new.eps}
\hspace{-5mm}
    \FigureFile(56mm,){fig6-9-new.eps}
  \end{center}
  \caption{
One-orbit (92-min) averages 
GSC light curves in 2-10~keV energy band and fitting results.
The dashed lines show the best-fit linear decay and 
the dash-dotted lines show the best-fit exponential decay.
Parameters are listed in table \ref{tab3}.
The dotted lines show the linear decay expected by brink model.
\textcolor{black}{
The region between two vertical lines is when there is 
a possibility of dip.}
}
\label{fig6}
\end{figure*}

\textcolor{black}{
In order to estimate the outer radius,
we chose nine outbursts with clear decay profiles in the 2--10~keV band
[(b), (g), (i), (j), (m), (p), (s), (t), and (u)]
from the remarkable eleven outbursts (in subsection 2.1)}.
Here we removed (c) and (l), as they showed a flat-top profile. 
First, we fitted 
\textcolor{black}{
the decay part by a linear function as}
\begin{equation}
L(\varphi) = - \gamma \times \varphi + C ,
\end{equation}
where $\varphi$ is the orbital phase, $\gamma$ is the slope, and
$C$ is the luminosity at $\varphi = 0$.
Here, $\varphi = 1$ is the next periastron after outburst onset.
The sudden drop occurs at $\varphi = 0.95$.
The fitting results are summarized in table~\ref{tab3} and 
indicated by dashed lines in figure~\ref{fig6}.

\begin{table*}
\caption{Parameters of linear and exponential decay fits of outbursts.}
\label{tab3}
\begin{center}
\begin{tabular}{ccccccccccc}
\hline
Outburst  & Interval \footnotemark[$*$] &
\multicolumn{4}{c}{Linear decay fit} & & \multicolumn{4}{c}{Exponential decay fit} \\
\cline{3-6} \cline{8-11} 
          &                             &
$\gamma$\footnotemark[$\dagger$] & $C$\footnotemark[$\ddagger$] &
$R_{\rm outer}$ \footnotemark[$\S$]& $\chi^{2}$ (dof) & &
$L_{\rm S}$\footnotemark[$\l$] & $\tau$\footnotemark[$\#$] &
$R_{\rm outer}$ \footnotemark[$**$]& $\chi^{2}$ (dof) \\
\hline
(b) & 0.30--1.5 \footnotemark[$\dagger\dagger$] &
$20.8^{+0.5}_{-0.4}$ & $33.1\pm0.6$ & $>$ 7.5 & 2.8 (133) &
& $32.9\pm0.8$ & $0.54^{+0.02}_{-0.01}$ & $4.82^{+0.09}_{-0.04}$ & 2.8 (132) \\
(g) & 0.30--1.9 \footnotemark[$\dagger\dagger$] &
$37.2^{+0.4}_{-0.5}$ & $70\pm1$ & $>$ 7.7 & 7.5 (202) &
& $64.5^{+0.7}_{-0.8}$ & $0.54^{+0.01}_{-0.01}$ & $4.82\pm0.04$  & 5.3 (201) \\
(i) & 0.46--0.95 & 
$83\pm6$ & $139\pm5$ & $>$ 7.2 & 7.1 (98) &
& $113\pm3$ & $0.88^{+0.06}_{-0.06}$ & $6.2^{+0.2}_{-0.3}$ & 6.9 (97) \\ 
(j) & 0.30--0.68 &
$29^{+9}_{-10}$ & $56\pm5$ & $>$ 8.4 & 1.1 (18) &
& $47^{+3}_{-2}$ & $1.4^{+0.7}_{-0.3}$ & $8^{+2}_{-1}$ & 1.1 (17) \\ 
(m) & 0.40--0.90 & 
$25\pm2$ & $39^{+2}_{-1}$ &  $>$ 7.0 & 2.5 (125) &
& $30.0^{+0.7}_{-1.0}$ & $0.90^{+0.08}_{-0.06}$ & $6.2^{+0.3}_{-0.2}$ & 2.5 (124) \\ 
(p) & 0.35--0.88 &
$46^{+8}_{-9}$ & $67^{+7}_{-6}$ & $>$ 6.9 & 2.1 (38) & 
& $55.5^{+4.8}_{-4.6}$ & $0.75^{+0.14}_{-0.10}$ & $5.7^{+0.5}_{-0.4}$ & 1.87 (37) \\
(s) & 0.44--0.92 &
$145^{+17}_{-34}$ & $160^{+21}_{-5}$ & $>$ 5.4 & 5.5 (21) & 
& $129^{+10}_{-9}$ & $0.37^{+0.05}_{-0.03}$ & $4.0^{+0.3}_{-0.2}$ & 4.01 (20) \\
(t) & 0.46--0.93 &
$130\pm6$ & $173\pm4$ & $>$ 6.1 & 2.6 (106) & 
& $119\pm2$ & $0.62^{+0.03}_{-0.02}$ & $5.2\pm0.1$ & 2.28 (105) \\
(u) & 0.22--0.93 &
$77^{+3}_{-1}$ & $96\pm2$ & $>$ 6.6 & 4.2 (171) &
& $88\pm2$ & $0.59^{+0.02}_{-0.02}$ & $5.0\pm0.1$ & 3.07 (170) \\
\hline
\multicolumn{11}{@{}l@{}}{\hbox to 0pt{\parbox{180mm}{\footnotesize
\par\noindent
\footnotemark[$*$]
Phase interval used for model fits.
\par\noindent
\footnotemark[$\dagger$]
Slope of linear decay in units of $10^{36}$ erg s$^{-1}$ phase$^{-1}$.
\par\noindent
\footnotemark[$\ddagger$]
Luminosity at phase 0 in units of $10^{36}$ erg s$^{-1}$.
\par\noindent
\footnotemark[$\S$]
Outer radius in units of $10^{10}$~cm of the accretion disk
estimated by the brink model,
assuming that the brink occurred at or before 
the beginning of the decay phase.
\par\noindent
\footnotemark[$\l$]
Luminosity at the beginning of the interval in units of $10^{36}$ erg s$^{-1}$.
\par\noindent
\footnotemark[$\#$]
Time constant of exponential decay in units of orbital period.
\par\noindent
\footnotemark[$**$]
Outer radius in units of $10^{10}$~cm of the accretion disk
estimated from the time constant of exponential decay.
\par\noindent
\footnotemark[$\dagger\dagger$]
Except for the phase of 0.95--1.3 to avoid dips. 
}\hss}}
\end{tabular}
\end{center}
\end{table*}

Next, we fitted each outburst by an exponential-decay function.
The X-ray luminosity is given by
\begin{equation}
L(\varphi) = L_{\rm S}\ {\rm exp} [-(\varphi - \varphi_{\rm S}) / \tau], 
\end{equation}
where $\varphi_{\rm S}$ is the phase at the beginning of the fitting,
which was fixed. 
$L_{\rm S}$ is the luminosity at $\varphi_{\rm S}$, 
and $\tau$ is the time-scale of the decay.
The fitting results are summarized in table~\ref{tab3}
and indicated by dash-dotted lines in figure~\ref{fig6}.

For outburst (b), \citet{DAi2012} reported that
the light curve is described well by a combination of two
linear decays with the 2--60~keV RXTE light curve for $\varphi$ = 0.3--1.68.
On the other hand, 
the GSC light curve in 2--10~keV can be fitted by either
a single linear decay or a single exponential decay
with the same reduced $\chi^{2}$.
The difference may be caused by the difference in intervals and 
sampling frequency between  the pointing observations with RXTE
and the monitoring  observations with MAXI. 

The outer disk radius $R_{\rm outer}$ can be estimated independently
using either the \textcolor{black}{
brink luminosity and the slope of linear decay (hereafter brink model)
}
or the time-scale of the exponential decay
(\cite{Powell2007}; \cite{Campana2013}).
In the brink model, using
the slope of the linear decay $\gamma$ and
the brink luminosity $L_{\rm brink}$,
$R_{\rm outer}$ can be derived as
\begin{equation}
R_{\rm outer} = (3\ \nu_{\rm KR}\ L_{\rm brink}\ /\ \gamma)^{1/2},
\label{equ}
\end{equation}
where $\nu_{\rm KR}$ is the viscosity at the edge of the outer disk.
King and Ritter (1998) adopted $\nu_{\rm KR} \sim 10^{15}$ cm$^{2}$ s$^{-1}$.
However, the brink feature was not recognized in all the outbursts.
Applying the \textcolor{black}{brink model}, we can assume that the brink
occurred at the beginning
of the decay phase or had already occurred earlier. 
Then the luminosity at the decay start $L_{\rm start}$
(at the phase corresponding to the first value
of the interval in table \ref{tab3}) 
is taken as the lower limit of the brink luminosity.
Then we can estimate the lower limit of
$R_{\rm outer}$ as
\begin{equation}
R_{\rm outer} > (3\ \nu_{\rm KR}\ L_{\rm start}\ /\ \gamma)^{1/2}.
\label{equ_start}
\end{equation}
The estimated outer disk radii are summarized in table~\ref{tab3}.

On the other hand, in the exponential decay model,
the outer disk radius is given by the time-scale $\tau$ of
the exponential decay as
\begin{equation}
R_{\rm outer} = (3\ \nu_{\rm KR}\ \tau)^{1/2} .
\end{equation}
The estimated outer disk radii are summarized in table~\ref{tab3}. 
The outer radii in the outbursts are estimated as $\sim 5\times10^{10}$ cm
assuming the exponential decay.
Here, we examined the relationship between the estimated outer radius
assuming exponential decay and the peak luminosity or
the luminosity just before the sudden drop around phase 
\textcolor{black}{
$\varphi= 0.95$.
}
There is no correlation between these luminosities and
the outer radius.

\section{Discussion}

We found sudden luminosity decreases \textcolor{black}{around} periastron
in 7 out of 21 outbursts observed with MAXI/GSC.
The luminosities dropped from a few times $10^{37}$ erg s$^{-1}$
to $\lesssim3\times10^{36}$ erg s$^{-1}$, and this was followed
by inactive states lasting for 14--16~d.
The light curve of the decay part can be fitted as well by either
a single linear function or a single exponential function.
In this section, we propose three interpretations
for the sudden luminosity decrease:
the end of the outburst during the dip, 
the propeller effect, and
the stripping effect by the stellar wind of the companion star.

\subsection{End of outburst during dip}

The sudden luminosity decrease in the seven outbursts,
(i), (j), (l), (m), (p), (s), and (u),
always occurred \textcolor{black}{during} periastron passage.
In Cir X-1, a dip was often observed \textcolor{black}{during}
periastron passage.
Therefore, we consider that
the luminosity became sufficiently low after the dip, 
and the outburst seems to have ended.
The dip was assumed to continue from phase 
\textcolor{black}{$\sim$}
0.95 to 
\textcolor{black}{$\sim$}
1.3
as observed in outburst (g)
(see figure~\ref{fig4}).
The decay profile during the dip was assumed to be three cases:
exponential decay, linear decay, and brink model.
In the case of exponential decay and linear decay,
we extrapolated the fitting results for the decay profile in subsection~2.3,
and plotted them in figure~\ref{fig6}.
The brink model assumes that the brink occurred during the dip, where
the decay profile changed from exponential to linear.
The slope $\gamma$ of the brink model can be limited by equation~(\ref{equ}).
In order to obtain the fastest decay case in the brink model,
the brink was assumed to occur at phase 0.95.
The $L_{\rm brink}$ values at phase 0.95 were obtained
by the exponential-decay fit.
The outer radius $R_{\rm outer}$ was calculated by the exponential-decay fit
as well.
Thus, we obtained the slope $\gamma$ of the brink model.
In figure~\ref{fig6}, the dotted lines show the linear decay expected from the
brink model.
Here, the outburst (l) is excluded since none of the exponential
and linear fittings worked for its flat-top profile (see subsection 2.3).

As shown in figure~\ref{fig6}, the observed data points after the drop lie
far below the three fitted lines except for outbursts (s) and (u).
This implies that it is difficult to explain the sudden drop in terms of the
interpretation that the outburst ended during the dip.
For outbursts (s) and (u), it is possible to explain the sudden drop
by a combination of the dip of phase 0.95--1.3 and
rapid linear decay, i.e., with linear decay or a brink in outburst (s),
and linear decay in outburst (u).

\subsection{Propeller Effect}

Sudden luminosity decrease is often attributed to the propeller effect
(Campana et al. 1998b, 2008; Matsuoka \& Asai 2013; Asai et al. 2013).
The propeller luminosity is related to the magnetic field and
spin period of the neutron star as \citep{Matsuoka2013}
\begin{equation}
L=1.5\times10^{37}\ \eta^{-7/2}P_1^{-7/3}B_8^2\ \rm{erg}~\rm{s}^{-1},
\end{equation}
where 
$P_1$ is the spin period of the neutron star in ms, and
$B_8$ is the magnetic field of the neutron star in units of $10^8$~G.
The values ($P_1$, $B_8$) are specific to the system, but 
they are not known for Cir~X-1.
Here $\eta \sim$ 0.5--1  is a factor representing the geometry of
the accretion flow;
it was introduced in the definition of the Alfv\'{e}n radius as
$R_{\rm A} = \ \eta\ R_{\rm A0}$, where $R_{\rm A0}$ is
the ideal Alfv\'{e}n radius in equation (27) in \citet{Gosh1979}.
\textcolor{black}{
The $\eta$ also appears in \citet{Burderi1998} as $\phi$.
}
Approximately,
$\eta \sim 0.5$ corresponds to a disk-like accretion flow and
$\eta \sim 1$ corresponds to a spherical accretion flow.
Thus, the propeller luminosity depends on the geometry of the accretion flow 
as $L \sim 1.3\times10^{36}\ P_1^{-7/3}B_8^2\ \rm{erg}~\rm{s}^{-1}$
if it is disk-like,
and $L \sim 1.5\times10^{37}\ P_1^{-7/3}B_8^2\ \rm{erg}~\rm{s}^{-1}$
if it is spherical.
\textcolor{black}{
Here, we notice that \citet{Wang1997} estimated
the $\eta$ ($\phi$ in Burderi et al. 1998) $\sim 1$
even in the case of disk-like accretion flow.
They calculated torque in inclined  magnetic moment axis to
the spin axis assuming that the spin axis is normal to the disk plane.
However, we use $\eta \sim$ 0.5--1 as was used traditionally.
}

The soft state is usually considered to be a manifestation of
an optically thick disk accretion.
The spectral states before the sudden drop are soft states
in all seven outbursts,
judging from the hardness ratio (BAT/GSC) in figure~\ref{fig2}.
\citet{DAi2012} reported that a soft-to-hard state transition occurred 
at a  luminosity of $(3.5\pm0.7)\times10^{36}$ erg s$^{-1}$
in the decay part of outburst (b).
This is within the commonly reported transition luminosity in the decay phase,
which is
1\%--4\% of the Eddington luminosity \citep{Maccarone2003}.
In outbursts (b) and (g) in figure~\ref{fig2},
the spectral state remained in the soft state above this luminosity,
which is consistent with \citet{DAi2012}.
\textcolor{black}{
This means
} 
there was no sudden drop in the soft-state period,
which indicates that the propeller effect did not occur above this luminosity.
Therefore, it is difficult to explain the sudden drop in terms of
the propeller effect.

Here we notice that the state-transition \textcolor{black}{may be}
still possible at high luminosity
in the truncated disk model of \citet{Gierlinski2008}.
If the inner disk is truncated at some point as the outburst decays,
the disk may recede.
Then, the disk-state  may transit from the soft state
to the hard state even at a luminosity higher than the usual
transition luminosity.
Then, the accretion flow may change from disk-like to spherical and
the propeller effect may occur during the dip,
\textcolor{black}{
because the propeller luminosity depends on the geometry of the 
accretion flow.}  
A careful examination of the behavior before the sudden drop
at phase
\textcolor{black}{
$\sim 0.95$
} in outbursts (i) and (j)
\textcolor{black}{
(see figure~\ref{fig5} and \ref{fig6})
}
reveals that the luminosity is quite variable, which might
suggest an unstable inner disk.

\subsection{Wind Stripping}

\begin{figure}
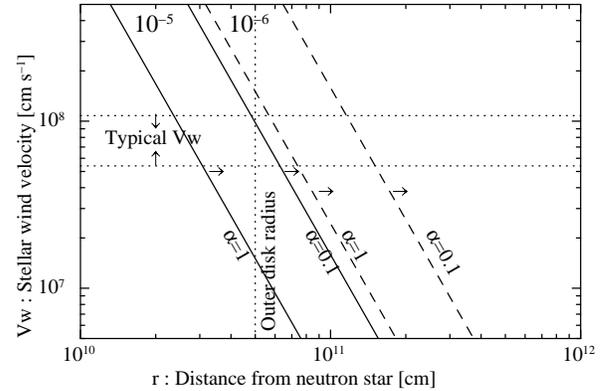

  \begin{center}
    \FigureFile(80mm,){fig7-new.eps}
  \end{center}
  \caption{
Allowed regions in a plane of stellar wind velocity and distance
from the neutron star.
Typical stellar wind velocity is between the horizontal dotted lines.
The vertical dotted line indicates estimated typical outer radius
assuming the exponential decay (see text).
The right side of solid and dashed lines are the region which fulfills
the condition of wind stripping.
\textcolor{black}
{
The left solid lines are
$\dot{M}_{\rm c} = 10^{-5}\Mo\rm{yr}^{-1}$
and the right dashed lines are
$\dot{M}_{\rm c} = 10^{-6}\Mo\rm{yr}^{-1}$.
A set of two lines indicate
when $\alpha = 1$ and 0.1.}
}
\label{fig7}
\end{figure}

A sudden drop can occur around periastron, where 
the stellar wind of the companion star interacts strongly with
the accretion disk around the neutron star
\citep{Johnston1999}.
If the stellar wind strips a large part of the accretion disk,
the disk may cause the state transition in the outer part.
As a result, mass accretion rate may decrease rapidly.

We examine the possibility of wind stripping.
The condition for wind stripping is
that the ram pressure of the stellar wind ($\rho_{\rm w} V_{\rm W}^2$)
is larger than the gas pressure of the accretion disk ($P_{\rm g}$).
Here, assuming a spherical stellar wind,
the density of the stellar wind is given by  
\begin{equation}
\rho_{\rm w} = \dot{M}_{\rm c} / (V_{\rm W}\ 4 \pi d^2),
\end{equation}
where $\dot{M}_{\rm c}$ is the mass loss rate of the companion star,
and $d$ is the distance from the center of the companion star.
The gas pressure of the accretion disk on the equatorial plane
is 
\begin{equation}
P_{\rm g} = G M_{\rm NS}~r^{-3}H^2\rho_{\rm d},
\end{equation}
where
$M_{\rm NS}$ is the mass of the neutron star,
$H$ is the half-thickness of the disk, and
$\rho_{\rm d}$ is the density on the equatorial plane of
the outer accretion disk.
The density, $\rho_{\rm d}$, is given by \citet{BHC-text} as
\begin{eqnarray}
\rho_{\rm d} = 37~\alpha^{-7/10}M_{1.4}^{-7/10}
(L_{\rm X}/L_{\rm E})^{11/20}  \nonumber \\ 
\times
(r/r_{\rm g})^{-15/8}f^{11/20}~\rm{g}~\rm{cm}^{-3},
\end{eqnarray}
where
$\alpha$ is the viscosity parameter,
$M_{1.4}$ is the neutron star mass normalized by $1.4M_\odot$,
$L_{\rm X}$ is the X-ray luminosity, 
$L_{\rm E}$ is the Eddington luminosity
($L_{\rm E} \sim 2 \times10^{38}$ erg s$^{-1}$),
$r$ is the distance from the neutron star,
$r_{\rm g}$ is the Schwarzschild radius of the neutron star,
and
$f = 1-(r_{\rm in}/r)^{1/2}$ 
(where $ r_{\rm in}$ is the inner disk radius).
The half thickness of the disk, $H$, is also given by \citet{BHC-text} as
\begin{eqnarray}
H = 2.0\times10^3~\alpha^{-1/10}M_{1.4}^{9/10}
(L_{\rm X}/L_{\rm E})^{3/20} \nonumber \\ 
\times
(r/r_{\rm g})^{9/8}f^{3/20}~\rm{cm}.
\end{eqnarray}

Figure~\ref{fig7} shows the allowed regions in the plane of
the stellar wind velocity and distance from the neutron star.
In this calculation,
we adopted $\dot{M}_{\rm c}$ = $10^{-5}\Mo\rm{yr}^{-1}$ and
$10^{-6}\Mo \rm{yr}^{-1}$.
\textcolor{black}{
The value of $d = 2.4\times10^{12}$~cm at periastron was calculated
using the third Kepler law and the values, 
eccentricity = 0.45, $M_{\rm c} = 10\Mo$, and orbital period = 16.68~d, 
in \citet{Jonker2007}.
}
We also adopted $\alpha$ =0.1--1, $L_{\rm X}=10^{37}$ erg s$^{-1}$,
and $f \sim 0.9$ ($r_{\rm in}/r \sim 1/100)$.
The typical stellar wind velocity is between the horizontal dotted lines.
\textcolor{black}{
Here, we adopted the typical value of
$V_{\rm W} = V_{\rm inf} \times (1-R/d)^\beta$,
where
$V_{\rm inf}$ is the terminal velocity; (1--2)$\times 10^8$~cm s$^{-1}$,
$\beta = 0.8$, and $R/d = 0.54$.
We adopted the radius of the companion star, $R$, as that of the Roche lobe
(=$1.3\times10^{12}$~cm) at periastron, although the radius is a little
smaller than that derived by spectral classification (B5--A0I)
\citep{Jonker2007}.   
The left solid lines are
$\dot{M}_{\rm c} = 10^{-5}\Mo\rm{yr}^{-1}$
and the right dashed lines are
$\dot{M}_{\rm c} = 10^{-6}\Mo\rm{yr}^{-1}$.
A set of two lines indicate
when $\alpha = 1$ and 0.1.
}
Here, the vertical dotted line indicates the estimated
typical outer disk radius
assuming the exponential decay model ($\sim 5\times10^{10}$cm).
To the right of solid and dashed lines,
the conditions for wind stripping are \textcolor{black}{fulfilled}.
\textcolor{black}{
Figure~\ref{fig7} shows that wind stripping works in the case of
$\dot{M}_{\rm c} = 10^{-5}\Mo\rm{yr}^{-1}$ and $\alpha = 1$.
This means that wind stripping occurs when the mass loss rate is large.
}

Next, we consider other conditions as follows.
For instance, the accretion radius could be larger than that derived from
exponential decay model and the ram pressure of the stellar wind is
likely about one order of magnitude higher than that of the
smooth wind adopted in this paper.
As shown in subsection~\ref{outerradius},
we cannot determine whether the decay is exponential or linear.
Here, we notice that periastron spikes do not occur
in the outbursts with a sudden drop
\textcolor{black}{
(see figure~\ref{fig1} and table~\ref{tab1})}.
This may indicate that the X-ray irradiation is smaller 
because there is
no spherical accretion onto the neutron star.
\textcolor{black}{
Then the decay would be more likely to be linear, and
the obtained outer radius of the accretion disk can be large
($\gtrsim 7 \times 10^{10}$ cm ), as shown in table~\ref{tab3}.
This larger radius is also compatible with the resonant radius,
(4--5) $\times 10^{11}$~cm,
which is calculated by the method of \citet{Artymowicz1994}.
}

Moreover, it is now widely accepted that the wind of OB stars is clumpy.
The core density becomes one order of magnitude higher than that
in the smooth wind
(e.g., \cite{Hamann2008}; \cite{Puls2008}).
With 10 times higher density wind in equation (8),
the solid and dashed lines
in figure~\ref{fig7} move to the left, and  
this change, together with the larger outer radius
($\gtrsim 7 \times 10^{10}\; \mathrm{cm}$), makes the
wind stripping model consistent with the observations.

Even when the ram pressure of the wind
does not exceed the gas pressure on the equatorial plane,   
it may blow out the surface of the accretion disk.
This may cause expansion of the disk followed by 
a decrease in $P_{\rm g}$. 
Then the interaction causes a state transition at the outer part of
the accretion disk.
The luminosity may suddenly drop below the soft-to-hard transition luminosity,
that is, $(3.5\pm0.7)\times10^{36}$ erg s$^{-1}$ \citep{DAi2012},
although 
\textcolor{black}{
it is very difficult to confirm}
that the state transition
occurred in the outer disk region in our results.

We also estimated the time scale of propagation of the sudden stripping
from the outer disk to the inner disk.
The viscous time scale is given by \citet{Frank1992} as 
\begin{equation}
t_{\rm visc}\sim 3 \times 10^5~\alpha^{-4/5}\,
\dot{M}_{16}^{-3/10} M_{1.4}^{1/4} \ R_{10}^{5/4}~\rm s,
\end{equation}
where
$\dot{M}_{16}$ is the mass accretion rate in units of $10^{16}$ g s$^{-1}$,
and 
$R_{10}$ is the typical length scale for surface density changes in
the disk in units $10^{10}$ cm.
\textcolor{black}{
Here, we assume} 
$\alpha$ = 0.1--1,
$\dot{M}_{16}$ = 5.3 ($L_{\rm x} \sim 10^{37}$~erg s$^{-1}$),
$M_{1.4}$ = 1,
and
$R_{10} \sim $
\textcolor{black}{
0.5 (the radius after wind stripping, which is one order of magnitude
smaller than the typical outer radius assuming the exponential decay),
and we find $t_{\rm visc}\sim$ 1--6~d.
The time scale is longer than that of sudden drop (a few hours).
However, if wind stripping occurs during the dip of phase 0.95--1.3,
the time scale is consistent with the observations.
}

\section{Conclusion}

Out of 21 outbursts observed with MAXI/GSC,
7 showed sudden luminosity decreases \textcolor{black}{around} periastron,
which led to the end of the outburst.
The light curve of the decay part can be fitted as well by either
a single linear function or a single exponential function.
We investigated three \textcolor{black}{interpretations} for
the sudden end of the outburst \textcolor{black}{around} periastron:
(1) the end of the outburst during the dip, 
(2) the propeller effect, and
(3) the stripping effect by the stellar wind of the companion star.
(1) is \textcolor{black}{possible for} only two outbursts, assuming linear decay.
(2) is not plausible because 
a luminosity significantly higher than 1\%--4\% of the Eddington luminosity
is required for a state transition.
If the inner disk is truncated, however, it is possible for the transition
to occur at a higher luminosity, and the sudden drop can be explained
by the propeller effect.
However, 
\textcolor{black}{
it is very difficult to confirm the} disk truncation in our results,
although the variability in the luminosity before the sudden drop
suggests instability in the inner disk.    
Finally, it is also difficult to explain the 
sudden drop by (3)
\textcolor{black}{
if a large mass loss rate is not assumed.
}
However, in the case of wind clumping and/or large outer radius assuming
linear decay, it is possible for the wind stripping to trigger the
sudden end of the outburst.
The interaction may cause the state transition at the outer part of
the accretion disk.
As a result, the luminosity may suddenly drop
below the soft-to-hard transition luminosity,
$(3.5\pm0.7)\times10^{36}$ erg s$^{-1}$.

\bigskip

We would like to acknowledge the MAXI team for MAXI operation
and for watching and analyzing real time data.
We also would like to acknowledge Toshihiro Takagi for his elaborate
analysis of milli-second pulsation search of Cir~X-1.
This research was partially supported by the Ministry of
Education, Culture, Sports, Science and Technology (MEXT),
Grant-in-Aid for Science Research 24340041.


\begin{thebibliography}{}

\bibitem[Armstrong et al.(2013)]{Armstrong2013}
Armstrong, R.~P., et al.\ 2013, \mnras, 433, 1951 

\bibitem[Artymowicz and Lubow(1994)]{Artymowicz1994}
\textcolor{black}{
Artymowicz, P., \& Lubow, S.~H.\ 1994, \apj, 421, 651 
}

\bibitem[Asai et al.(2012)]{Asai2012}
Asai, K., et al.\ 2012, \pasj, 64, 128 

\bibitem[Asai et al.(2013)]{Asai2013}
Asai, K., et al.\ 2013, \apj, 773, 117 

\bibitem[Barthelmy et al.(2005)]{Barthelmy2005}
Barthelmy,~S.~D., et al. 2005, \ssr, 120, 143

\bibitem[Boutloukos et al.(2006)]{Boutloukos2006}
Boutloukos, S., van der Klis, M., Altamirano, D., Klein-Wolt, M., 
Wijnands, R., Jonker, P. G., \& Fender, R. P. 2006, \apj, 653, 1435 

\bibitem[Burderi et al.(1998)]{Burderi1998}
Burderi, L., di Salvo, T., Robba, N.~R., del Sordo, S., 
Santangelo, A., \& Segreto, A. 1998, \apj, 498, 831 

\bibitem[Brandt et al.(1996)]{Brandt1996}
Brandt, W.~N., Fabian, A.~C., Dotani, T., Nagase, F., Inoue, H.,
Kotani, T. \& Segawa, Y. \ 1996, \mnras, 283, 1071 

\bibitem[Bradt et al.(1993)]{Bradt1993}
  Bradt,~H.~V., Rothschild,~R.~E., \& Swank,~J.~H. 1993, \aaps, 97, 355

\bibitem[Brandt \& Schulz(2000)]{Brandt2000}
Brandt, W.~N., \& Schulz, N.~S.\ 2000, \apj, 544, L123 

\bibitem[Campana et al.(1998a)]{Campana1998a}
Campana,~S., Colpi,~M., Mereghetti,~S., Stella,~L., \& Tavani,~M. 1998
\aapr, 8, 279

\bibitem[Campana et al.(2013)]{Campana2013}
Campana, S., Coti Zelati, F., \& D'Avanzo, P.\ 2013, \mnras, 432, 1695 

\bibitem[Campana et al.(2008)]{Campana2008}
Campana, S., Stella, L., \& Kennea, J.~A.\ 2008, \apj, 684, L99 

\bibitem[Campana et al.(1998b)]{Campana1998b}
Campana, S., Stella, L., Mereghetti, S, Colpi, M., Tavani, M., 
Ricci, D., Fiume, D. Dal, \& Belloni, T.\ 1998, \apj, 499, L65 

\bibitem[Clarkson et al.(2004)]{Clarkson2004}
Clarkson, W.~I., Charles, P.~A., \& Onyett, N.\ 2004, \mnras, 348, 458 

\bibitem[D'A{\`i} et al.(2012)]{DAi2012}
D'A{\`i}, A., et al.\ 2012, \aap, 543, 20 


\bibitem[Frank et al.(1992)]{Frank1992}
Frank, J., King, A., \& Raine, D.\ 1992,
Accretion power in astrophysics.
(Camb.~Astrophys.~Ser., Vol.~21)

\bibitem[Gehrels et al.(2004)]{Gehrels2004}
Gehrels,~N., et al. 2004, \apj, 611, 1005

\bibitem[Gierli{\'n}ski et al.(2008)]{Gierlinski2008}
Gierli{\'n}ski, M., Done, C., \& Page, K.\ 2008, \mnras, 388, 753 

\bibitem[Ghosh and Lamb(1979)]{Gosh1979}
Ghosh, P., \& Lamb, F.~K.\ 1979, \apj, 232, 259 

\bibitem[Hamann et al.(2008)]{Hamann2008}
Hamann, W.-R., Feldmeier, A., \& Oskinova, L.~M.\ 2008,
Clumping in Hot-Star Winds


\bibitem[Heinz et al.(2013)]{Heinz2013}
Heinz,~S., et al. 2013, \apj, 779, 171

\bibitem[Iaria et al.(2001a)]{Iaria2001a}
Iaria, R., Burderi, L., Di Salvo, T., La Barbera, A., \& Robba, N.~R
\ 2001a, \apj, 547, 412 

\bibitem[Iaria et al.(2001b)]{Iaria2001b}
\textcolor{black}{
Iaria, R., Di Salvo, T., Burderi, L., \& Robba, N.~R.\ 2001b, \apj, 561, 321 
}

\bibitem[Jonker et al.(2007)]{Jonker2007}
Jonker, P.~G., Nelemans, G., \& Bassa, C.~G.\ 2007, \mnras, 374, 999 

\bibitem[Johnston et al.(1999)]{Johnston1999}
Johnston, H.~M., Fender, R., \& Wu, K.\ 1999, \mnras, 308, 415 

\bibitem[Kaluzienski et al.(1976)]{Kaluzienski1976}
Kaluzienski, L.~J., Holt, S.~S., Boldt, E.~A., \& Serlemitsos, P.~J.\ 1976, \apjl, 208, L71 

\bibitem[Kato et al.(1998)]{BHC-text}
Kato, S., Fukue, J., \& Mineshige, S.\ 1998,
Black-hole accretion disks.~ Edited by Shoji Kato, Jun Fukue,
and Sin Mineshige.~ Publisher: Kyoto, Japan: Kyoto University Press,
1998.~ISBN: 4876980535

\bibitem[King \& Ritter(1998)]{King1998}
King, A.~R., \& Ritter, H.\ 1998, \mnras, 293, L42 

\bibitem[Kirsch et al.(2005)]{Kirsch2005}
Kirsch,~M.~G.,et al. 2005, Proc. SPIE, 5898, 22

\bibitem[Lasota(2001)]{Lasota2001}
Lasota, J.-P.\ 2001, New Astronomy Reviews, 45, 449 

\bibitem[Levine et al.(1996)]{Levine1996}
Levine, A.~M.,  et al.\ 1996, \apjl, 469, L33 

\bibitem[Lin et al.(2012)]{Lin2012}
Lin, D., Remillard, R.~A., Homan, J., \& Barret, D.\ 2012, \apj, 756, 34 


\bibitem[Linares et al.(2010)]{Linares2010}
Linares, M., et al.\ 2010, \apjl, 719, L84 

\bibitem[Matsuoka et al.(2009)]{Matsuoka2009}
Matsuoka,~M., et al. 2009, \pasj, 61, 999

\bibitem[Matsuoka \& Asai(2013)]{Matsuoka2013}
Matsuoka, M., \& Asai, K.\ 2013, \pasj, 65, 26 

\bibitem[Maccarone(2003)]{Maccarone2003}
Maccarone,~T.~J. 2003, \aap, 409, 697

\bibitem[Mihara et al.(2011)]{Mihara2011}
Mihara,~T., et al. 2011, \pasj, 63, S623

\bibitem[Mohamed \& Podsiadlowski(2007)]{Mohamed2007}
Mohamed, S., \& Podsiadlowski, P.\ 2007,
15th European Workshop on White Dwarfs, 372, 397 

\bibitem[Nakajima et al.(2010)]{Nakajima2010}
Nakajima, M., et al.\ 2010, The Astronomer's Telegram, 2608, 

\bibitem[Negoro et al.(2010)]{Negoro2010}
 Negoro,~H., et al. 2010, ASPC, 434, 127N

\bibitem[Nicolson(2007)]{Nicolson2007}
Nicolson, G.~D.\ 2007, The Astronomer's Telegram, 985, 1

\bibitem[Parkinson et al.(2003)]{Parkinson2003}
Parkinson, P.~M.~S., et al. 2003, \apj, 595, 333 

\bibitem[Powell et al.(2007)]{Powell2007}
Powell, C.~R., Haswell, C.~A., \& Falanga, M.\ 2007, \mnras, 374, 466 

\bibitem[Puls et al.(2008)]{Puls2008}
Puls, J., Vink, J.~S., \& Najarro, F.\ 2008, \aapr, 16, 209 

\bibitem[Shirey et al.(1999a)]{Shirey1999a}
Shirey, R.~E., Bradt, H.~V., \& Levine, A.~M.\ 1999a, \apj, 517, 472 

\bibitem[Shirey et al.(1999b)]{Shirey1999b} 
Shirey, R.~E., Levine, 
A.~M., \& Bradt, H.~V.\ 1999b, \apj, 524, 1048 

\bibitem[Soleri et al.(2009)]{Soleri2009}
Soleri, P., Tudose, V., Fender, R., van der Klis, M.,
\& Jonker, P.~G.\ 2009, \mnras, 399, 453 

\bibitem[Sugizaki et al.(2011)]{Sugizaki2011}
Sugizaki,~M., et al. 2011, \pasj, 63, S635

\bibitem[Tennant et al.(1986)]{Tennant1986}
Tennant, A.~F., Fabian, A.~C., \& Shafer, R.~A.\ 1986, \mnras, 221, 27

\bibitem[Tomida et al.(2011)]{Tomida2011}
Tomida,~H., et al. 2011, \pasj, 63, 397

\bibitem[Wang(1997)]{Wang1997} Wang, Y.-M.\ 1997, \apj, L475, 
L135 

\bibitem[White(1989)]{White1989} White, N.~E.\ 1989, \aapr, 1, 85 

\end{thebibliography}
\end{document}